\documentclass[twocolumn,aps,prl,groupedaddress,showpacs]{revtex4-1}
\usepackage{amsfonts}
\usepackage{amsmath}
\usepackage{amssymb}
\usepackage{txfonts}
\usepackage{pxfonts}
\usepackage{graphicx,bm,units,yfonts}
\usepackage[table]{xcolor}
\usepackage{helvet}         
\usepackage{courier}        
\usepackage{makeidx}

\makeindex

\begin{document}

\title{Virtual Topological Insulators with Real Quantized Physics}
\author{Emil Prodan}
\affiliation{Department of Physics, Yeshiva University, New York, NY 10016, USA}

\begin{abstract}
A concrete strategy is presented for generating strong topological insulators in $d+d'$ dimensions which have quantized physics in $d$ dimensions. Here, $d$ counts the physical and $d'$ the virtual dimensions. It consists of seeking $d$-dimensional representations of operator algebras which are usually defined in $d+d'$ dimensions where topological elements display strong topological invariants. The invariants are shown, however, to be fully determined by the physical dimensions, in the sense that their measurement can be done at fixed virtual coordinates. We solve the bulk-boundary correspondence and show that the boundary invariants are also fully determined by the physical coordinates. We analyze the virtual Chern insulator in $(1+1)$-dimensions realized in Ref.~\cite{KrausPRL2012hh} and predict quantized forces at the edges. We generate a novel topological system in $(3+1)$-dimensions, which is predicted to have quantized magneto-electric response.  
\end{abstract}


\maketitle

\section{Introduction}

In Refs.~\cite{KrausPRL2012hh,KrausPRL2013fj},  Kraus {\it et al} put forward a framework in which topological insulators can be generated in arbitrary dimensions but with some of these dimensions occurring in a parameter space. For example, a Chern insulator defined in one physical dimension and one virtual dimension was proposed and realized in the laboratory using coupled one-dimensional wave guides, and several interesting topological effects were pointed out. The paper states at some point that the entire approach could work in any number of dimensions, and indeed, a Chern insulator of the second kind in $2+2$ dimensions was found in  Ref.~\cite{KrausPRL2013fj}. The virtual Chern insulator in $1+1$ dimensions was since observed in several other experiments \cite{VerbinPRL2013kj,VerbinPRB2015vc,HuPRX2015gf}. 

There is certainly a common feature between the $1+1$- and $2+2$-dimensional systems mentioned above, namely the quasi-crystalline structure, but beyond that they seem to be exceptional isolated topological examples of this type. One interesting question is if these systems actually fit into a larger class unified by a common principle? The defining and unifying property was already stated in \cite{KrausPRL2012hh} for $(1+1)$ dimensions, and could be generalized as follows: A virtual $(d+d')$-dimensional topological insulator is characterized by a strong topological invariant in $(d+d')$-dimensions but the invariant is measurable or computable entirely from the physical $d$ dimensions. How can such systems exist? As we shall see, the reason is two-fold: 
\begin{enumerate}
\item Operator algebras which are naturally defined or represented in higher dimensions sometime accept faithful ({\it i.e.} no information is lost) representations in lower dimensions. 
\item A certain self-averaging property occurs in these lower dimensions, which has the effect of integrating out the virtual dimensions. 
\end{enumerate}

These principles were already stated in \cite{KrausPRL2012hh,KrausPRL2013fj}, but since there was no explicit expression of the invariant, there was some confusion and, for example, the claims were challenged in \cite{MadsenPRB2013bd}. In their response letter \cite{KrausArxiv2013}, the authors reiterated the statements  but again no expression for the invariants was provided. They conceded in this letter that the boundary phenomena depend on the virtual dimensions, alluding that the topological class cannot be determined from  boundary phenomena without exploring the virtual dimensions. In this work, we use an operator algebra formalism to provide the missing expressions of the topological invariants. We pin-point the fundamental difference between the irrational and rational cases, which is the ergodic vs non-ergodic character of the lattice translations. The self-averaging property, which exists only for the irrational case, can then be rigorously formulated using Birkhoff's ergodic theorem \cite{BirkhoffPNAS1931jf}. We also solve the bulk-edge correspondence and prove that the boundary invariant, which is equal to the bulk invariant,  is entirely computable from the physical dimensions. In other words, the topological class can be determined even if we have access only to the boundary of the systems. The formalism also enables us to make three new physical predictions for the experimentalists. 

We recall that the bulk-edge correspondence principle is not just about the emergence of the boundary states.  The bulk-boundary program was defined and solved by Hatsugai  \cite{HatsugaiPRB1993cs} for the clean integer quantum Hall effect, and consists of 1) defining a boundary invariant, 2) proving that the boundary and bulk invariants are equal, 3) determining the physical meaning of the boundary invariant. In the general setting, which includes disorder and irrational magnetic flux values, the bulk-boundary problem was solved by Kellendonk, Richter and Schulz-Baldes \cite{SchulzBaldes:2000p599,Kellendonk:2002of} using a $K$-theoretic approach. Our analysis relies on these works, which are adapted to the present context without detailed proofs. This is mainly because  a comprehensive manuscript on the bulk-boundary principe for the complex classes is in preparation \cite{ProdanInPrep}. For the virtual Chern insulator in $(1+1)$-dimensions, we find that the boundary invariant is equal to the average mechanical force spontaneously exerted at the edge of the system. The existence of such quantized forces were first predicted in Refs.~\cite{KellendonkJPA2004ve,JPAKellendonk2005dt}. We also formulate and analyze the boundary invariant for the virtual second-Chern insulator in $3+1$ dimensions.

 The analysis is carried out at the level of operator algebras, which we find useful for the following reasons:
\begin{itemize}

\item It provides a natural framework for the analysis of homogeneous aperiodic systems \cite{BellissardLN2003bv}. It enables us to incorporate incommensurate and random potentials yet keep the calculations explicit. As such, we are able to translate all the claims made in \cite{KrausPRL2012hh} into rigorous mathematical statements and to strengthen some of them, especially those referring to the edge physics. While we agree that in condensed matter physics the ultimate confirmation comes from experiment, the mathematics may be appreciated here because it gives definitive answers to some of the open issues. We also feel that we can teach the reader in these 13 pages how to slightly tweak the old tools and adapt them to these more general settings. 

\item It provides the big picture, in the sense that seemingly unrelated models can be connected by one relatively simple operator algebra. Many operator algebras have been studied independently and a great deal of information is available (see for example \cite{DavidsonBook1996bv}). Perhaps the most important piece of information is the $K_0$-group which encodes the topology of the projections, in particular, of the spectral projections of the Hamiltonians generated from the algebra. Then the patterns seen in the energy spectrum, such as the Hofstadter butterfly, can be qualitatively and quantitatively understood from the $K_0$-group alone \cite{BellissardLNP1986jf}. The $K_0$-group of the rotational algebra, which generates our example of $(3+1)$-dimensional topological insulator, has 8 generators among which one has non-zero second-Chern number (equal to 1). As such, even if we generate the Hamiltonians randomly, we have 12.5\% chance to generate a second-Chern insulators, hence no numerics are needed to prove that such Hamiltonians exist.  

\item Lastly, it provides an outstanding computational toolbox. Note that no proof (or reference) was provided in \cite{KrausPRL2012hh,KrausPRL2013fj} for the quantization of the invariants in the strict irrational settings. After all, the whole point \cite{KrausArxiv2013} there was that one cannot ``break the symmetry" of the quasi-crystal and go into the rational setting where one can use a range of arguments. The index theorems for the irrational settings, which give the quantization of the first \cite{BELLISSARD:1994xj} and higher \cite{ProdanJPA2013hg}  Chern numbers, were established using the operator algebra formalism. Also, for example, it is well known that one cannot use perturbation theory even for small magnetic fields, hence computing response functions involving the derivatives with respect to the magnetic field can be a challenging task (see for example \cite{EssinPRB2010ls} in the context of magneto-electric response of topological insulators). With operator algebras, one has a well defined calculus \cite{RammalJPF1990ki}, that is, a set of explicit rules which reduce these calculations to a simple exercise in differential geometry (see for example \cite{LeungJPA2013er} for a computation of the magneto-electric response function done in this way, and to be used here). For the systems defined in \cite{KrausPRL2012hh}, these tools can be applied to investigate the behavior of various quantities with respect to the parameter $b$ \cite{KrausPRL2012hh} (called $\theta$ here). In particular, one can show that the invariants are continuous of $b$ (hence constant) if a gap happens to remain open, a fact which was assumed in \cite{MadsenPRB2013bd} but questioned in \cite{KrausArxiv2013}, for example. 
\end{itemize}

\section{A Virtual Chern Insulator in 1+1-dimensions}

\subsection{The System Defined}

The first example of a virtual topological insulator consists of a 1-dimensional periodic chain subjected to an incommensurate onsite potential:
\begin{equation}\label{HPhi}
H_\phi = T_1 + T_1^\dagger + 2W \cos( \theta X+ \phi),
\end{equation}
where $T_n$ represents the translation operator by $n$ unit cells $T_n|x\rangle = |x+n\rangle$, and $X$ the position operator $X|x\rangle = x|x\rangle$. The Hamiltonian has a localization-delocalization transition at the critical value $W_c=1$: For $W<1$, the entire spectrum is delocalized  and for $W>1$ the spectrum is localized \cite{AvilaLNP2006yt}. Throughout, we will set the fundamental constants to one.

This is the topological model considered in the Ref.~\cite{KrausPRL2012hh} and further analyzed in a number of subsequent works \cite{KrausPRL2012ds,LangPRL2012ew,VerbinPRL2013kj,MadsenPRB2013bd,LiuPRB2015tr}. While most of the related experimental works have been focused on photonics, here we adopt a different view and think of this model as that of a molecular chain absorbed on a substrate with incommensurate lattice. The angle $\theta$ quantifies the substrate's lattice constant relative to that of the chain and the angle $\phi$ quantifies how the molecular structure is aligned with respect to the substrate's lattice.  In a real experiment, one deals with an entire ensemble of absorbed molecular chains, for which $\phi$ is expected to take random values in the interval $[0,2\pi]$. 

\begin{figure}
\includegraphics[width=8.6cm]{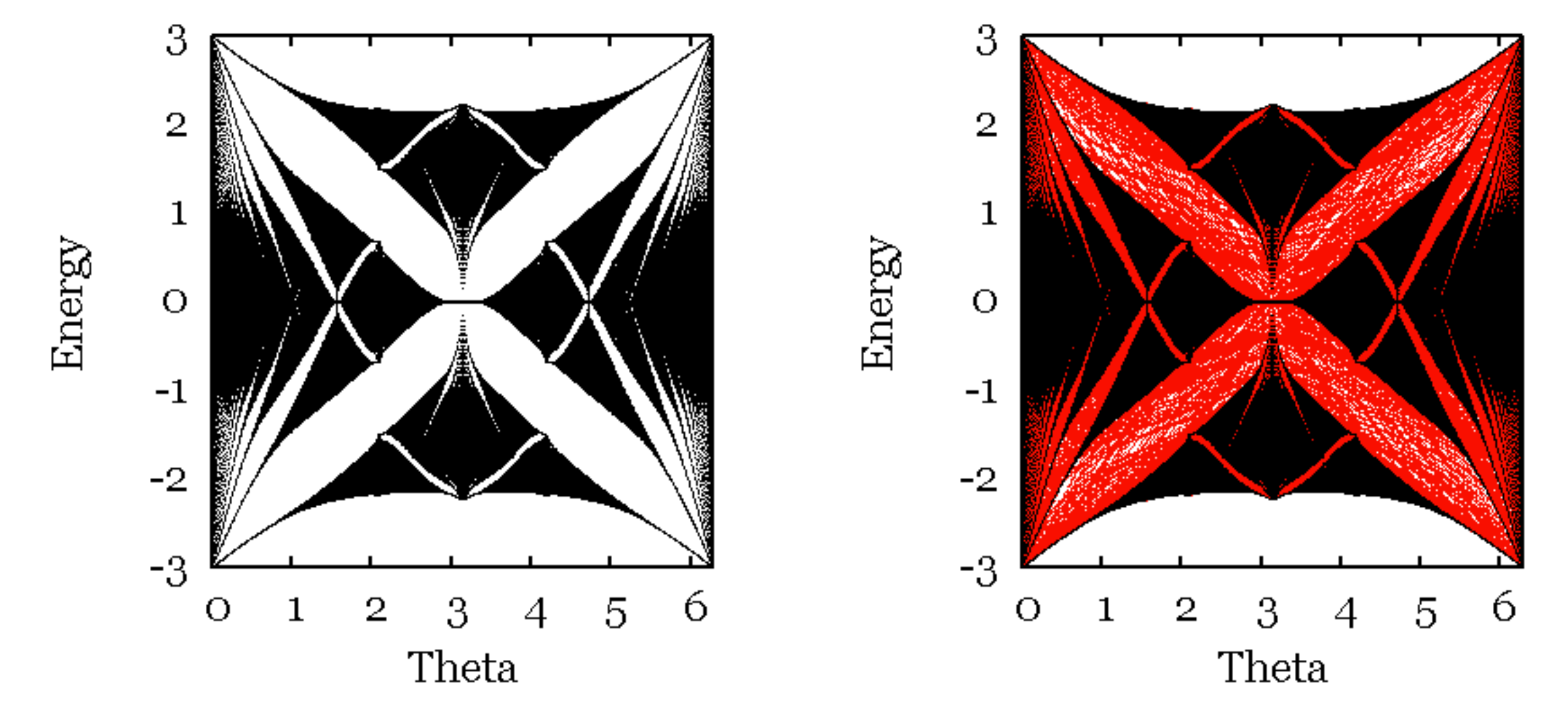}
\caption{(Color online) Left panel: The energy spectrum of $H_\phi$ defined in Eq.~\ref{HPhi} as function of $\theta$, computed with periodic boundary conditions on a finite 1 dimensional lattice of $10^3$ sites. Right panel: The energy spectrum of $H_\phi$ as function of $\theta$, computed on a 400-sites lattice but with open boundary conditions. The plot shows a superposition of 100 spectra obtained with 100 random values of $\phi$ in the interval $[0,2\pi]$. For guidance, the bulk spectrum from the left panel is shown with darker color. The value of $W$ is subcritical, $W=0.5$.}
\end{figure} 

The bulk energy spectrum of $H_\phi$ as function of $\theta$ is shown in the left panels of Figs.~1 and 2 for subcritical and overcritical values $W=0.5$ and $W=1.5$, respectively. The spectra are identical with the corresponding Hofstadter spectra (see discussion below). The energy spectrum of an ensemble of long but finite chains with open boundary conditions and random $\phi$-parameter is shown in the right panels of Figs.~1 and 2. They reveal that, regardless of the localized or delocalized character of the bulk states, almost all bulk energy gaps are completely filled with edge spectrum. These are indeed the trades of a topological insulator.

\subsection{Connection with the Hofstadter Model}

Since $H_\phi$ is nothing but the Harper or the almost-Mathieu Hamiltonian, its connection with models describing two dimensional electrons in magnetic fields is well known. Here we provide an algebraic connection which will help us shape the principle for constructing virtual topological insulators. Quite surprisingly, a key role is played by the angle $\phi$. 

Note first that under translations:
\begin{equation}
T_n H_\phi T_n^\dagger = T_1 + T_1^\dagger + 2W \cos( \theta X+ \phi+n \, \theta), 
\end{equation}
or:
\begin{equation}
T_n H_\phi T_n^\dagger = H_{(\phi+n \, \theta){\rm mod} \, 2\pi}.
\end{equation}
As a result, the translations generate a dynamical system $(\mathbb S,\tau)$ on the unit circle $\mathbb S$:  \begin{equation}
\mathbb S \ni \phi \rightarrow \tau \phi = (\phi+\theta)\, {\rm mod} \ 2\pi.
\end{equation} 
Acting repeatedly with $\tau$, we obtain an action of $\mathbb Z$ on $\mathbb S$:
\begin{equation}
\mathbb Z \ni n \rightarrow \tau_n \phi = \stackrel{n \ {\rm times}}{\overbrace{\tau\circ \ldots \circ \tau}}\phi = (\phi+n \, \theta) \, {\rm mod} \ 2\pi.
\end{equation}
As we shall see, this dynamical system is key to the self-averaging property mentioned in the introduction.

\begin{figure}
\includegraphics[width=8.6cm]{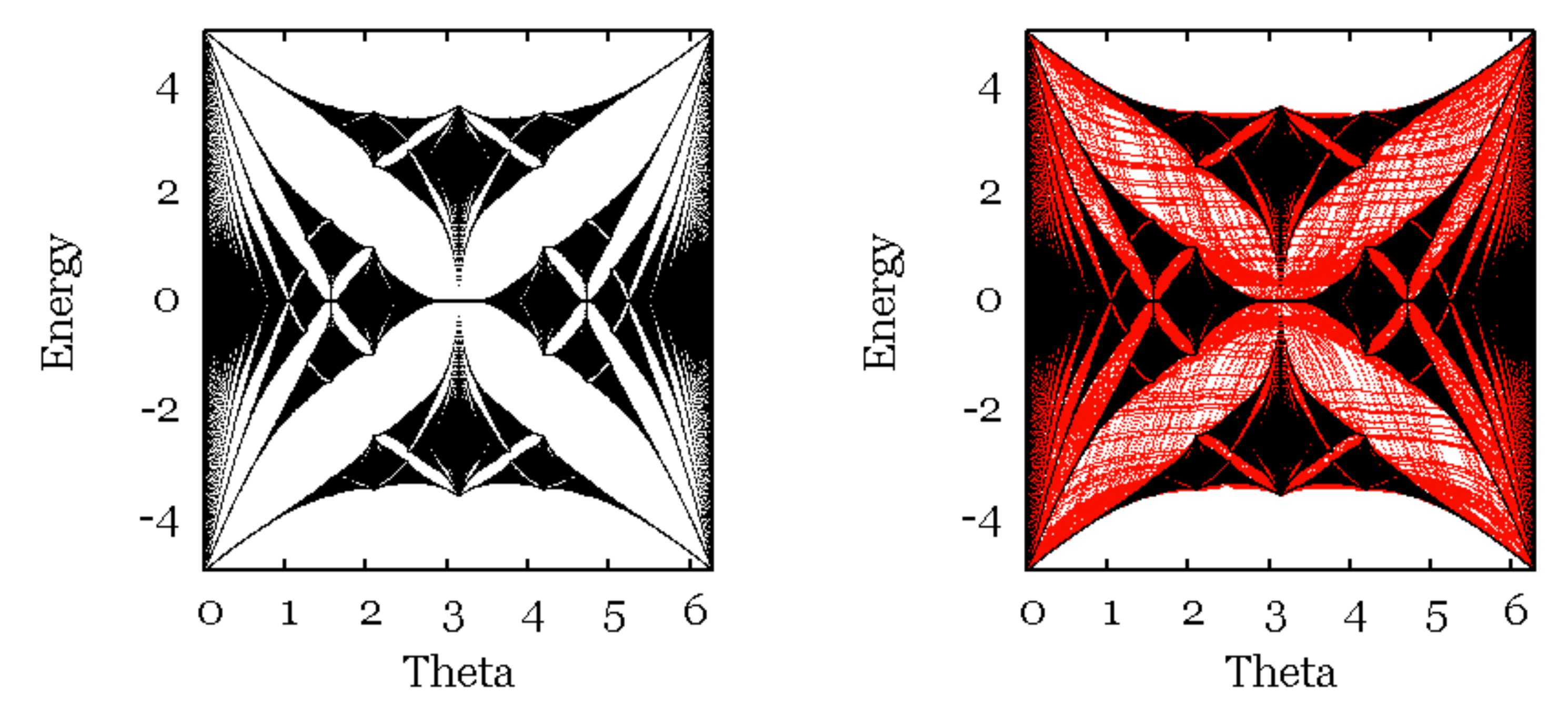}
\caption{(Color online) Same as Fig.~1 but for the overcritical value $W=1.5$.}
\end{figure}

Next, we define a dual dynamical system. For this, one considers the algebra $C(\mathbb S)$ of complex valued continuous functions over the unit circle. $\tau_n$ implements an action of the group $\mathbb Z$ on $C(\mathbb S)$ in the following way:
\begin{equation}\label{ZAction1}
C(\mathbb S) \ni f \rightarrow \alpha_n(f) = f \circ \tau_n.
\end{equation}
Pointwise, this is equivalent to:
\begin{equation}
f(\phi) \overset{\alpha_n}{\longrightarrow} f(\tau_n \phi), \ \forall \phi \in [0,2\pi].
\end{equation} 
The dual dynamical system is $\big ( C(\mathbb S),\alpha \big )$. In the theory of operator algebras, there is a standard construction which enlarges the algebra $C(\mathbb S)$ so that the action defined in Eq.~\ref{ZAction1} is implemented by unitary internal elements. It is called the crossed product of $C(\mathbb S)$ by $\mathbb Z$ and is denoted by $C(\mathbb S) \rtimes_\alpha \mathbb Z$. It is obtained by augmenting a unitary operator $u$ to $C(\mathbb S)$ and enforcing the commutation relation:
\begin{equation}\label{Commutation1}
u^n f u^{-n} = \alpha_n(f), \ \forall \ f \in C(\mathbb S).
\end{equation}
Then the elements of $C(\mathbb S) \rtimes_\alpha \mathbb Z$ are formal series:
\begin{equation}\label{GenericElement1}
\bm a = \sum_{n\in \mathbb Z} f_n u^n, \ \  f_n \in C(\mathbb S).
\end{equation}
If $\bm b = \sum_{n\in \mathbb Z} g_n u^n$ is another element, then the commutation relation in Eq.~\ref{Commutation1} leads to:
\begin{equation}
\bm a \bm b = \sum_{n\in \mathbb Z} \big ( \sum_{m \in \mathbb Z} f_m \ \alpha_{m}(g_{n-m}) \big ) u^n,
\end{equation}
which should be seen as a convolution that is twisted by the map $\alpha$. We should mention that the crossed product algebra $C(\mathbb S) \rtimes_\alpha \mathbb Z$ is one of the most studied algebras in operator algebra theory \cite{RieffelPJM1981bf}.

The algebra $C(\mathbb S) \rtimes_\alpha \mathbb Z$ accepts a family of canonical representations on the Hilbert space $\ell^2(\mathbb Z)$ of square-summable functions over $\mathbb Z$:
\begin{equation}
\pi_\phi(\bm a)  = \sum_{n,x \in \mathbb Z} f_n(\tau_x \phi) \, |x \rangle \langle x | \, T_n , \ \ \phi \in [0,2\pi].
\end{equation}
It is a simple matter of term counting for one to convince himself that indeed $\pi_\phi(\bm a) \pi_\phi(\bm b) = \pi_\phi(\bm a \bm b)$. The family of Hamiltonians defined in Eq.~\ref{HPhi} are representations
\begin{equation}
H_\phi = \pi_\phi(\bm h)
\end{equation} 
of the following element:
\begin{equation}\label{HElement}
\bm h = u + u^\ast + 2 W \cos(\phi).
\end{equation}

We now consider the Hofstadter Hamiltonian on the 2-dimensional lattice $\mathbb Z^2$:
\begin{equation}\label{Hofstadter}
H_\Phi = \mathcal T_{(1,0)} + \mathcal T_{(1,0)}^\dagger + W \big (\mathcal T_{(0,1)} + \mathcal T_{(0,1)}^\dagger \big)
\end{equation}
where:
\begin{equation}
\mathcal T_{(n,m)} |x,y\rangle = e^{- i  m x \Phi } |x+n,y+m \rangle
\end{equation} 
are the dual magnetic translations and obey the following commutation relation:
\begin{equation}
\mathcal T_{(1,0)} \mathcal T_{(0,1)} = e^{i \Phi} \mathcal T_{(0,1)}\mathcal T_{(1,0)}.
\end{equation}
In this form, the Hofstadter Hamiltonian is invariant with respect to the magnetic translations:
\begin{equation}
\mathcal T'_{(n,m)} |x,y\rangle = e^{i  n y \Phi } |x+n,y+m \rangle,
\end{equation}
written here in the Landau gauge. Above, $\Phi$ represents the magnetic flux through the unit cell, expressed in units $\hbar /e$.

We show in the following that $H_\Phi$ is in fact a representation on $\ell^2(\mathbb Z^2)$ of the same element $\bm h$ from Eq.~\ref{HElement}. Indeed, since any continuous function over $\mathbb S$ can be decomposed as a discrete Fourier series,  one can see from Eq.~\ref{GenericElement1} that the algebra $C(\mathbb S) \rtimes_\alpha \mathbb Z$ is generated by $z=e^{i \phi}$ and $u$. The commutation relation between these generators is:
\begin{equation}
uz = e^{i\theta} zu,
\end{equation}
which is exactly the commutation relation between the $\mathcal T$'s, if we enforce $\Phi=\theta$. If that is the case, we immediately obtain a representation of $C(\mathbb S) \rtimes_\alpha \mathbb Z$ on $\ell(\mathbb Z^2)$, by sending $z$ into $\mathcal T_{(1,0)}$ and $u$ into $\mathcal T_{(0,1)}$. In other words:
\begin{equation}
\pi'(\bm a) = \sum_{n \in \mathbb Z} f_n \big (\mathcal T_{(1,0)} \big ) \big ( \mathcal T_{(0,1)} \big )^n.
\end{equation}
It is evident that $H_\Phi = \pi'(\bm h)$ with the same $\bm h$ as in Eq.~\ref{HElement}.

\subsection{Analysis of the Virtual Topological Insulator}

Given the connection between $H_\phi$ and $H_\Phi$, we can understand at once why their energy spectra are identical. Indeed, recall that the abstract element $\bm h$ itself posses a spectrum, defined by those $\xi \in \mathbb C$ for which $\bm h -\xi \cdot \bm 1$ is not invertible. If $\theta$ is irrational, then the representations $\pi$ and $\pi'$ are faithful, and consequently the spectra of $\bm h$, $\pi_\phi(\bm h)$ and $\pi'(\bm h)$ coincide. We also know from Ref.~\cite{BellissardCMP1994gf} that the gap edges in the Hofstadter spectrum are continuous with respect to $\theta$, hence the spectra coincide also for the rational values of $\theta$. One should be aware that this statement concerns strictly the location of the spectrum and that it says nothing about the localized or delocalized nature of the spectrum.

The next thing we want to investigate is the nature of the virtual dimension and how to compute the bulk topological invariant. Using the generators $z$ and $u$, we can write the generic elements as:
\begin{equation}\label{MNRep}
\bm a = \sum_{m,n \in \mathbb Z} c_{mn} z^m u^n.
\end{equation}
This representation reveals the 2-dimensional nature of the problem, as $(m,n)$ live on the 2-dimensional lattice $\mathbb Z^2$. The coordinate $n$ is related to the real-space coordinate $x$ but the coordinate $m$ is abstract and relates to the Fourier decomposition of the $f_n$ functions. For example, the generator $\bm h$ of the Hamiltonian $H_\phi$ is give by:
\begin{equation}
\bm h = u + u^\ast + W(z + z^\ast).
\end{equation}

We now follow the work of Bellissard on aperiodic solids \cite{BellissardLNP1986jf} and introduce a non-commutative calculus on $C(\mathbb S) \rtimes_\alpha \mathbb Z$. It is defined by a formal (but mathematically rigorous) integration $\mathcal I$:
\begin{equation}\label{Integration1}
\mathcal I (\bm a) = c_{00},
\end{equation} 
and two derivations:
\begin{equation}
\partial_1 \bm a = i\sum_{m,n \in \mathbb Z} m \, c_{mn} z^m u^n
\end{equation}
and
\begin{equation}
\partial_2 \bm a = i\sum_{m,n \in \mathbb Z} n \, c_{mn} z^m u^n.
\end{equation}
In the representation given in Eq.~\ref{GenericElement1}, the first derivation is just the ordinary derivation with respect to $\phi$:
\begin{equation}
\partial_1 \bm a = \sum_{n \in \mathbb Z} (\partial_\phi f_n) \, u^n,
\end{equation} 
and
\begin{equation}\label{Integration11}
\mathcal I (\bm a) = \tfrac{1}{2\pi}\int_{\mathbb S}d \phi \ f_0(\phi).
\end{equation}
Together, $\big ( C(\mathbb S) \rtimes_\alpha \mathbb Z, \bm \partial, \mathcal I \big )$ define a non-commutative manifold dubbed \cite{BellissardLNP1986jf} the non-commutative Brillouin torus.

Now let $\chi(t) = \frac{1}{2}(1 + {\rm sgn}(t))$ be the usual step function on the real axis and consider the Fermi projection:
\begin{equation}
\bm p= \chi (\epsilon_F -\bm h),
\end{equation}
 that is, the spectral projection of $\bm h$ onto the spectrum below the Fermi level $\epsilon_F$. We define the bulk invariant as the 1-st non-commutative Chern number \cite{BELLISSARD:1994xj}:
\begin{equation}\label{2DChern1}
{\rm Ch}_1(\bm p) = 2 \pi i \ \mathcal I \big ( \bm p [\partial_1 \bm p, \partial_2 \bm p]\big ),
\end{equation}
which is a topological invariant \cite{BELLISSARD:1994xj} in the sense that, as long as $\epsilon_F$ is located in a spectral gap of $\bm h$, ${\rm Ch}_1(\bm p) $ takes integer values and it remains constant under continuous deformations of $\bm h$.

As already elaborated in the introduction, one of the main statements of \cite{KrausPRL2012hh} was that the bulk invariant can be computed at fixed $\phi$. The implication of this statement is that the topology is encoded in a single physical system $H_\phi$ rather than in its whole family $\{H_\phi\}_{\phi \in \mathbb S}$. These statements were reinforced in \cite{KrausArxiv2013}, in response to the apparently contradicting conclusions in \cite{MadsenPRB2013bd}. Now, examining Eq.~\ref{2DChern1} and the definition of the integration in Eq.~\ref{Integration11}, there seems to be a contradiction here. But contrary, we show that the statement of \cite{KrausPRL2012hh} is correct and can be formulated in mathematically rigorous terms. To start, let us use Birkhoff's ergodic theorem \cite{BirkhoffPNAS1931jf} to rewrite the integration as:
\begin{equation}
\mathcal I (\bm a) = \tfrac{1}{2\pi}\int_{\mathbb S}d \phi \ f_0(\phi)= \lim_{N\rightarrow \infty} \frac{1}{2N}\sum_{|x|\leq N} f_0(\tau_x \phi).
\end{equation}
Then we use the representation $\pi_\phi$ to write:
\begin{equation}
f_0(\tau_x \phi) = \langle x | \pi_\phi (\bm a)|x \rangle
\end{equation}
and now one can see that the integration is nothing but the trace per length ${\rm Tr}_L$ of a physical representation at {\bf fixed} $\phi$:
\begin{equation}
\mathcal I (\bm a) = \lim_{N\rightarrow \infty} \frac{1}{2N}\sum_{|x|\leq N}\langle x | \pi_\phi (\bm a)|x \rangle =  {\rm Tr}_L \Big (\pi_\phi (\bm a) \Big ).
\end{equation}
Furthermore, using the simple rules: $\pi_\phi (\partial_1 \bm a)= \partial_\phi \pi_\phi(\bm a)$ and $\pi_\phi(\partial_2 \bm a) = i[X,\pi_\phi(\bm a)]$, we obtain the desired expression for the bulk invariant:
\begin{equation}
{\rm Ch}_1(\bm p) = - 2 \pi \ {\rm Tr}_L \Big ( P_\phi \big [\partial_\phi  P_\phi, [X, P_\phi] \big ] \Big ),
\end{equation}
where $P_\phi = \chi(\epsilon_F - H_\phi)$ is the physical Fermi projection. The statement in \cite{KrausPRL2012hh} is confirmed. A fine point which was made in \cite{KrausPRL2012hh,KrausArxiv2013} was that the statement is true only if $\theta$ is irrational. We now can see explicitly the reason behind this statement: the map $\tau_x$ is ergodic on $\mathbb S$ for $\theta$ irrational hence Birkhoff's ergodic theorem applies (w.r.t. the usual measure), but this is not the case if $\theta$ is rational. The former case can be visualized by imagining the sequence $\phi, \tau \phi, \tau^2 \phi, \ldots$ being inked on a piece of paper, in which case one will discover that, no matter how sharp the writing instrument, the unit circle will always be entirely covered  by ink. This is obviously not the case if $\theta$ is rational, in which case the sequence of dots will closed into itself and there will only be a finite number of marks. Hence, the system is a virtual topological insulator, as defined in the introduction, only if $\theta$ is irrational.

\subsection{Robustness Against Disorder}

The disorder can occur in the chain's lattice or in the substrate. In a tight-binding approach, the disorder introduces random fluctuations in all the coefficients of the Hamiltonian:
\begin{align}\label{HPhiOmega}
H_{\bm \omega,\phi} &= \sum_{x \in \mathbb Z}(1+\lambda \omega_x) \big ( |x\rangle \langle x+1| +|x \rangle \langle x -1| \big) \\
 & + 2W \sum_{x \in \mathbb Z} (1+\lambda ' \omega'_x)\cos( x \, \theta + \phi) \, |x \rangle \langle x|. \nonumber
\end{align}
The distributions of $\omega_x$ and $\omega'_x$ should be determined from the energetics of the lattice distortions, but here we will assume that they are independently and randomly generated from the interval $[-\frac{1}{2},\frac{1}{2}]$. One should note that the original Hamiltonian is recovered when $\lambda$ and $\lambda'$ are set to zero.

As it follows from the works of Bellissard \cite{BellissardLN2003bv}, the disorder can be treated algebraically too. For this, one considers the disorder configuration space 
\begin{equation}
\Omega = [-\tfrac{1}{2},\tfrac{1}{2}]^{\mathbb Z} \times [-\tfrac{1}{2},\tfrac{1}{2}]^{\mathbb Z},
\end{equation}
 such that $\bm \omega = (\omega,\omega') \in \Omega$ gives at once all the random amplitudes in the Hamiltonian: 
\begin{equation}
\Omega \ni \bm \omega = \{\omega_x ,\omega'_x \}_{x \in \mathbb Z}.
\end{equation}
The probability measure to be used for disorder averaging is simply 
\begin{equation}
d\omega = \prod_x d\omega_x d\omega'_x.
\end{equation}
 Next, one defines the ergodic dynamical system $(\Omega \times \mathbb S, \tau)$, with:
\begin{align}
& \tau_n(\bm \omega,\phi)  = (\tau_n \bm \omega,\tau_n \phi) \nonumber \\
&= \big (\{\omega_{x+n}, \omega'_{x+n}\}_{x \in \mathbb Z}, (\phi+n \, \theta) \, {\rm mod} \, 2\pi \big ).
\end{align}
The relevance of this dynamical system comes from the fact that:
\begin{equation}
T_n H_{\bm \omega,\phi} T_n^\dagger = H_{\tau_n \bm \omega,\tau_n \phi}.
\end{equation}
Systems possessing such covariance property are called homogeneous.

The dual dynamical system $\big ( C(\Omega \times \mathbb S),\alpha \big )$ is defined by:
\begin{equation}
\alpha_n(f) = f \circ \tau_n, \ \forall \ f\in C(\Omega \times \mathbb S).
\end{equation}
Pointwise, this is equivalent to:
\begin{equation}
f(\bm \omega,\phi) \overset{\alpha_n}{\longrightarrow} f(\tau_n \bm \omega, \tau_n \phi), \ \forall \ (\bm \omega,\phi) \in \Omega \times \mathbb S.
\end{equation}
As before, one can form the crossed product algebra $C(\Omega \times \mathbb S) \rtimes_\alpha \mathbb Z$ by augmenting a unitary element $u$ to the algebra of continuous functions over $\Omega \times \mathbb S$, such that:
\begin{equation}\label{Commutation2}
u^n \, f \, u^{-n} = \alpha_n (f).
\end{equation} 
The elements of this algebra are formal series of the form:
\begin{equation}\label{GenericElement2}
\bm a = \sum_{n\in \mathbb Z} f_n u^n, \ \  f_n \in C(\Omega \times \mathbb S),
\end{equation}
and, if $\bm b = \sum_{n\in \mathbb Z} g_n u^n$ is another element, then the commutation relation in Eq.~\ref{Commutation2} leads to:
\begin{equation}\label{MultiRule}
\bm a \bm b = \sum_{n\in \mathbb Z} \big ( \sum_{m \in \mathbb Z} f_m \ \alpha_{m}(g_{n-m}) \big ) u^n.
\end{equation}
This looks formally the same as Eq.~\ref{Commutation1}, but note that $f_n$ and $\alpha_n$ have a new meaning.

The algebra $C(\Omega \times \mathbb S) \rtimes_\alpha \mathbb Z$ accepts a family of canonical representations:
\begin{equation}
\pi_{\bm \omega,\phi}(\bm a)  = \sum_{n,x \in \mathbb Z} f_n(\tau_x \bm \omega,\tau_x \phi) \, |x \rangle \langle x | \, T_n.
\end{equation}
and it is easy to see that the disordered Hamiltonian in Eq.~\ref{HPhiOmega} is the representation
\begin{equation}
H_{\bm \omega,\phi} = \pi_{\bm \omega,\phi}(\bm h')
\end{equation}
 of the following element:
\begin{equation}\label{HElement2}
\bm h' = (1+\lambda \omega_0)(u + u^\ast) + 2W(1+\lambda' \omega'_0) \cos(\phi).
\end{equation}   
Here and throughout, by $\omega_0$ we mean the function which assigns to $\omega = \{\omega_x\}_{x \in \mathbb Z}$ the real number $\omega_0$. Note that when $h'$ is multiplied by another element or, for example, is squared, the result of the multiplication (see Eq.~\ref{MultiRule}) will contain $\omega_{\tau_x0}=\omega_x$, explicitly.

Following again the work on IQHE of Bellissard and his collaborators \cite{BELLISSARD:1994xj}, we define a non-commutative calculus over the new algebra. First, let us note that the elements of this algebra can be represented as in Eq.~\ref{MNRep}, but this time the coefficients $c_{mn}$ are not just numbers but functions over $\Omega$:
\begin{equation}\label{MNOmegaRep}
\bm a = \sum_{m,n \in \mathbb Z} c_{n,m}(\bm \omega) \, z^m u^n.
\end{equation}
The noncommutative Brillouin torus is defined as the non-commutative space $$\big ( C(\Omega \times \mathbb S) \rtimes_\alpha \mathbb Z, \bm \partial, \mathcal I \big )$$ where the derivations remain the same as before but the integral is modified as:
\begin{equation}
\mathcal I(\bm a) = \int_\Omega d \bm \omega \ c_{00}(\bm \omega). 
\end{equation}
In the representation given in Eq.~\ref{GenericElement2}:
\begin{equation}
\mathcal I (\bm a) = \tfrac{1}{2\pi}\int_{\mathbb S}d \phi \int_\Omega d \bm \omega \, \ f_0(\bm \omega,\phi).
\end{equation}

Consider now the Fermi projection:
\begin{equation}
\bm p'= \chi (\epsilon_F -\bm h') = \sum_{m,n \in \mathbb Z} p'_{mn}(\bm \omega) \, z^m u^n.
\end{equation}
According to Ref.~\cite{BELLISSARD:1994xj}, as long as:
\begin{equation}\label{Condition2}
\sum_{m,n \in \mathbb Z^2} (m^2+n^2) \int_\Omega d \bm \omega \ |p'_{mn}(\bm \omega)| < \infty,
\end{equation}
the first Chern number:
\begin{equation}\label{2DChern2}
{\rm Ch}_1(\bm p') = 2 \pi i \ \mathcal I \big ( \bm p' [\partial_1 \bm p', \partial_2 \bm p']\big )
\end{equation}
remains quantized and constant under continuous deformations of $\bm h'$. This invariant can be evaluated numerically using the elementary methods developed in Refs.~\cite{ProdanPRL2010ew,ProdanJPhysA2011xk,ProdanAMRX2013bn}. For example, even in the regime where the Fermi level is imbedded in dense localized spectrum, the quantization of the Chern numbers can be obtained with machine precision when these algorithms are use. 

To understand the physical meaning of the mathematical condition~\ref{Condition2}, first note that:
\begin{equation}
\pi_{\bm \omega,\phi} (\bm p')  = \chi(\epsilon_F - H_{\bm \omega,\phi}) =P_{\omega,\phi},
\end{equation}
{\it i.e.} the Fermi projection for one physical representation. Then:
\begin{equation}
p'_n(\bm \omega, \phi) = \langle 0 |P_{\omega,\phi}|n \rangle,
\end{equation}
and ondition~\ref{Condition2} can now be translated into:
\begin{equation}\label{Co1}
 \sum_{n\in \mathbb Z} n^2 \int\limits_{\mathbb S} d\phi \int\limits_\Omega d\bm \omega \ \big |\langle 0 |P_{\omega,\phi}|n \rangle\big|^2 < \infty
\end{equation}
and 
\begin{equation}\label{Co2}
\sum_{n\in \mathbb Z}  \int\limits_{\mathbb S} d\phi \int\limits_\Omega d\bm \omega \  \big | \partial_\phi \langle 0 |P_{\omega,\phi}|n \rangle\big|^2 < \infty.
\end{equation}
These two conditions must be {\it simultaneously} satisfied for the quantization and homotopy invariance of the first Chern number to hold. While Eq.~\ref{Co1} is automatically satisfied if the Fermi level resides in a region of Anderson-localized spectrum, this may not be the case for Eq.~\ref{Co2}. However, the later condition is for sure satisfied if the Fermi level resides in a clean spectral gap.  

Since the dynamical system $(\Omega \times \mathbb S, \tau)$ is ergodic, we can follow exactly the same arguments as before to equivalently expressed the bulk invariant as:
\begin{equation}
{\rm Ch}_1(\bm p') = - 2 \pi \ {\rm Tr}_L \Big ( P_{\omega,\phi} \big [\partial_\phi  P_{\omega,\phi}, [X, P_{\omega,\phi}] \big ] \Big ).
\end{equation}
As one can see, both integrals over $\phi$ and $\omega$ disappear, which tells us that the invariant can be computed at fixed virtual coordinate $\phi$ and from one disorder configuration. 

\subsection{Physical Prediction for Bulk}

The physical meaning of the invariant can be determined by adapting the non-commutative Kubo-formula \cite{BELLISSARD:1994xj,Schulz-Baldes:1998vm,Schulz-Baldes:1998oq} to the present context. This leads us to the prediction that by moving the chain relative to the substrate, hence varying the angle $\phi$ in time, one can set a charge current $J$ along the chain. The time-average value of the current is:
\begin{equation}
J = {\rm Ch}_1(\bm p') \ \dot \phi .
\end{equation}
This is our physical prediction, concerning the bulk. This effect will be present in the absence of edges, such as in a ring geometry.

\subsection{Bulk-Boundary Correspondence}

We now consider the half-space system by restricting the physical space from $\mathbb Z$ to $\mathbb N$. Consequently, we replace the Hilbert space $\ell^2(\mathbb Z)$ with $\ell^2(\mathbb N)$. As it follows from the works of Kellendonk, Richter and Schulz-Baldes \cite{SchulzBaldes:2000p599,Kellendonk:2002of}, this new physics can also be treated algebraically. To understand the required modifications, let us enquire about the faith of the translation operator, after we project onto the half of the space:
\begin{equation}
T_1 \rightarrow \hat T_1 = T_1 \Pi,
\end{equation}
where $\Pi$ is the projection from $\ell^2(\mathbb Z)$ to $\ell^2(\mathbb N)$.  We have:
\begin{equation}
\hat T_1|x\rangle = |x+1 \rangle, \ \forall \ x\in \mathbb N,
\end{equation} 
and 
\begin{equation}
\hat T_1^\dagger \, |x\rangle = |x-1 \rangle, \ \forall \ x>0, \ {\rm and} \  \hat T_1^\dagger \, |0\rangle=0.
\end{equation}
As a consequence, the translation is no longer a unitary operator and instead is a partial isometry:
\begin{equation}
\hat T_1^\dagger \hat T_1 = I, \ \ \hat T_1 \hat T_1^\dagger = I - |0\rangle \langle 0|.
\end{equation}
This suggests that the only modification we need to make to the algebra $C(\Omega \times \mathbb S) \rtimes_\alpha \mathbb Z$ is to replace the unitary operator $u$ by a partial isometry $\hat u$:
\begin{equation}\label{Isometry1}
\hat u^* \hat u =1, \ \ \hat u \hat u^\ast =1 -\hat e,
\end{equation}
where $\hat e$ is a projection, $\hat e^2=\hat e$, $\hat e^\ast=\hat e$. This projection relates to $|0 \rangle \langle 0|$ once a physical representation is considered. The elements of the new algebra, denoted by $C(\Omega \times \mathbb S) \rtimes_\alpha \mathbb N$ (not a standard notation), are formal series:
\begin{equation}\label{GenericElement3}
\hat{\bm a} = \sum_{m,n\in \mathbb N} f_{mn} \, \hat u^m (\hat u^\ast)^n, \ \  f_{mn} \in C(\Omega \times \mathbb S).
\end{equation}

Now, any element $\bm a = \sum_{n\in \mathbb Z} f_n \, u^n$ from the bulk algebra $C(\Omega \times \mathbb S) \rtimes_\alpha \mathbb Z$ can be transformed into an element from the half-space algebra:
\begin{equation}
\bm a \rightarrow j(\bm a) = \sum_{n\in \mathbb Z} f_n \, \hat{u}^n,
\end{equation}
where by $\hat u^{-n}$ it is understood $(\hat u^\ast)^n$. Note that $\hat u$ is not invertible hence $\hat u^{-1}$ will be incorrect to use. Also note that $j$ is only a linear map and not an algebra homomorphism (it does not respect the multiplication rule). Reciprocally, any element $\hat{\bm a} = \sum_{m,n\in \mathbb N} f_{mn} \, \hat u^m (\hat u^\ast)^n$ from the half-space algebra can be transformed into an element from the bulk-algebra:
\begin{equation}
\hat{\bm a} \rightarrow {\rm ev}(\hat{\bm a}) = \sum_{m,n\in \mathbb N} f_{mn}  \,  u^{m-n},
\end{equation}
by sending $\hat u$ into $u$ and $\hat u^\ast$ into $u^{-1}$. This time, the map ${\rm ev}$ respects both the addition and multiplication operations, hence it is an homomorphism of algebras.  

Using Eqs.~\ref{Isometry1}, it follows immediately that any element from $C(\Omega \times \mathbb S) \rtimes_\alpha \mathbb N$ can be written uniquely as:
\begin{equation}
\hat{\bm a} = j(\bm a) + \tilde{\bm a},
\end{equation}
with
\begin{equation}
\bm a = {\rm ev}(\hat{\bm a}) \ \ {\rm and} \ \ \tilde{\bm a} = \hat{\bm a}- j(\bm a).
\end{equation}
The component $\tilde{\bm a}$ is necessarily of the form:
\begin{equation}
\tilde{\bm a} = \sum_{m,n\in \mathbb N} \tilde f_{mn} \, \hat u^m \, \hat e \, (\hat u^\ast)^n, \ \  \tilde f_{mn} \in C(\Omega \times \mathbb S),
\end{equation}
and the elements like $\tilde{\bm a}$ form an ideal inside the half-space algebra, which we call the edge-algebra. A formal integral can be defined over the edge algebra:
\begin{equation}
\widetilde{\mathcal I}(\tilde{\bm a}) = \sum_{n}\int_{\mathbb S} d\phi \int_\Omega d\bm \omega \ \tilde f_{nn}(\bm \omega, \phi).
\end{equation}
 
The elements of the half-space algebra accept a canonical representation as operators over $\ell^2(\mathbb N)$:
\begin{equation}\label{Rep3}
\begin{array}{c}
 \hat{\bm a} = j(\bm a) + \tilde{\bm a} \rightarrow   \hat \pi_{\bm \omega, \phi} (\hat{\bm a} ) \medskip \\
 = \Pi \pi_{\bm \omega,\phi}(\bm a)\Pi + \sum\limits_{m,n\in \mathbb N} \tilde f_{mn}(\tau_m\bm \omega,\tau_m \phi) \hat T_m |0 \rangle \langle 0|\hat T_n^\dagger,
\end{array}
\end{equation}
where $\pi_{\bm \omega,\phi}(\bm a)$ is the bulk representation already discussed above. We can see that the first term represents the restriction of the bulk operator $\pi_{\bm \omega,\phi}(a)$ to the half-space via the open boundary condition, while the second term represents the component of the half-space operator that is localized at the edge. For example, the family of half-space Hamiltonians $\Pi H_{\bm \omega,\phi} \Pi$ is generated by:
\begin{equation}
j(\bm h') =(1+\lambda \omega_0)(\hat u + \hat u^\ast) + 2W(1+\lambda' \omega'_0) \cos(\phi).
\end{equation}
The open boundary condition can be changed to any other boundary condition by adding an element from the edge algebra:
\begin{equation}
\tilde{\bm h}= \sum_{m,n\in \mathbb N} \tilde h_{mn} \, \hat u^m \, \hat e \, (\hat u^\ast)^n.
\end{equation}
This term can account for the lattice distortions or relaxations near the edge, chemical contamination, bond breaking and so on. The statements below are true for any such generic boundary condition.

We now follow again Kellendonk, Richter and Schulz-Baldes \cite{SchulzBaldes:2000p599,Kellendonk:2002of}, and define the topological invariant for the edge states. We will consider the general case with disorder and an arbitrary boundary condition at the edge. A bulk spectral gap is assumed. Hence, let $\widehat H_{\bm \omega,\phi}$ be a family of half-space Hamiltonians, generated by $\hat{\bm h}'$ from the half-space algebra:
\begin{equation}
\hat{\bm h}' = j(\bm h') + \tilde{\bm h}',
\end{equation}
with $\bm h'$ from the bulk algebra and given in Eq.~\ref{HElement2}, and $\tilde{\bm h}'$ a generic element from the edge algebra. The construction of the edge invariant starts from the bulk Fermi projection $\bm p'= \chi(\epsilon_F - \bm h')$. The first task is to find an element $\hat{\bm g}$ from the half-space algebra such that ${\rm ev}(\hat{\bm g}')=\bm p'$. In mathematical terms, we are searching for a lift of $\bm p'$ from the bulk to the half-space algebra. In our context, this means:
\begin{equation}
\hat{\bm g}' = j(\bm p') + \tilde{\bm p}'.
\end{equation}
The solution is quite simple. Consider a smooth version $\tilde \chi$ of the step function $\chi$, such that $\tilde \chi(\epsilon)=1/0$ above/below a small interval around the origin (hence the entire smooth variation happens inside the small interval). Then $\hat{\bm g}'$ can be taken as:
\begin{equation}
\hat{\bm g}' = \tilde \chi\big (\epsilon_F - \hat{\bm h}' \big )
\end{equation}
because, since the difference between $\chi$ and $\tilde \chi$ occurs inside the bulk energy gap, 
\begin{equation}
\bm p'= \chi(\epsilon_F - \bm h')= \tilde \chi(\epsilon_F - \bm h').
\end{equation} 
and one can write:
\begin{equation}
\tilde{\bm p}'=\tilde \chi(\epsilon_F - \hat{\bm h'}) - j \big (\tilde \chi(\epsilon_F - \bm h') \big ),
\end{equation}
which finally can be shown to be localized near the edge, hence belongs to the edge algebra. This will not be the case if $\tilde \chi$ were replaced by $\chi$. In other words, the smoothness of $\tilde \chi$ is essential in this construction.

The next step is to consider the unitary element:
\begin{equation}\label{UnitaryF}
\hat{\bm u}'= \exp \big ( -2 \pi i \hat{\bm g}' \big )
\end{equation} 
in the half-space algebra. Then the edge invariant is simply the winding number of $\hat {\bm u}'$ in one dimension:
\begin{equation}\label{EdgeInvariant1}
\nu_1(\hat{\bm u}') = - \widetilde{\mathcal I}\big ( \hat{\bm u}' \partial_\phi (\hat{\bm u}')^\ast \big ).
\end{equation}
The first fundamental result of Ref.~\cite{Kellendonk:2002of} is the equality between the bulk and edge invariants:
\begin{equation}
{\rm Ch}_1(\bm p') = \nu_1(\hat{\bm u}').
\end{equation}

Let us now consider the physical representation:
\begin{equation}
U_{\bm \omega, \phi} = \hat \pi_{\bm \omega,\phi} (\hat{\bm u}'),
\end{equation}
which can be computed directly from:
\begin{equation}
\widehat U_{\bm \omega, \phi} = \exp \Big ( -2 \pi i \tilde \chi \big (\epsilon_F - \widehat H_{\bm \omega,\phi} \big ) \Big )
\end{equation}
by diagonalizing $\widehat H_{\bm \omega, \phi}$ or by any other methods of functional calculus. Then the invariant takes a more familiar form:
\begin{equation}\label{EdgeInvariant2}
\nu_1(\hat{\bm u}') = - \tfrac{1}{2\pi} \int_{\mathbb S} d \phi \int_\Omega d \bm \omega \ {\rm Tr}\big \{\widehat U_{\bm \omega,\phi} \partial_\phi \widehat U_{\bm \omega, \phi}^\dagger \big \}.
\end{equation}
This is a true edge invariant because $\widehat U_{\bm \omega,\phi}$ can be replace by $\widehat U_{\bm \omega,\phi} - I$ in \ref{EdgeInvariant2}. Looking back at the construction, it immediately becomes apparent that $\widehat U_{\bm \omega,\phi}-I$ can be built entirely from the spectrum and the states inside the bulk energy gap, {\it i.e.} from the boundary states. One can see at once that, if the bulk invariant is not zero, then the bulk spectral gap must be filled entirely with edge spectrum when $\phi$ is scanned over the interval $[0,2\pi]$. Indeed, if there was a gap left, then we can place the Fermi level inside this gap in which case $\widehat U_{\bm \omega,\phi} - I=0$ and consequently $\nu_1(\hat{\bm u}')=0$. But this will contradict the equality between the bulk and boundary invariants, hence no spectral gap can occur in the edge spectrum. Note that this statement is now established for any boundary condition.

We now show that the boundary invariant, hence also the bulk invariant, can be retrieved from the boundary physics at a fixed $\phi$ and disorder configuration. An interesting remark in \cite{KrausArxiv2013} was that if the boundary of one physical realization is etched atom by atom, then boundary states will appear and disappear in the process, and if we mark the energies of all these states, then these marks will eventually fill the insulating gap entirely. This will reveal indeed the topological character but it is not clear how one will recover the bulk invariant by just observing the edge eigenvalues (think of some large value, ${\rm Ch}_1 = 10$, for example). Now, when the atoms are etched, the boundary is moved but we can shift the whole system (chain + substrate) and place the boundary back in the original position. This set of moves will effectively shift the disorder and advance the phase $\phi$ by $\theta$. Now consider the quantity ${\rm Tr}\{(\widehat U_{\bm \omega,\phi}-I) \partial_\phi \widehat U_{\bm \omega, \phi}^\dagger\}$, which can be computed entirely from the boundary states. If we average the proposed quantity as the boundary is etched atom by atom, then we can see again Birkhoff's ergodic theorem in action:
\begin{align}
&-\lim_{N\rightarrow \infty} \frac{1}{N} \sum_{x<N}  {\rm Tr}\big \{(\widehat U_{\tau_x \bm \omega,\tau_x \phi}-I) \partial_\phi \widehat U_{\tau_x \bm \omega, \tau_x \phi}^\dagger \big \}  \\
&  = - \tfrac{1}{2\pi} \int_{\mathbb S} d \phi \int_\Omega d \bm \omega \ {\rm Tr}\big \{(\widehat U_{\bm \omega,\phi}-I) \partial_\phi \widehat U_{\bm \omega, \phi}^\dagger \big \},
\end{align} 
and the boundary invariant emerges. The statement is proved.

\subsection{Physical Interpretation and Predictions}

Returning to physics, let us state the second fundamental result of Refs.~\cite{SchulzBaldes:2000p599,Kellendonk:2002of}, which is the following equivalent formula for the edge invariant:
\begin{equation}\label{EdgeInvariant3}
\nu_1 (\hat{\bm u}') = - \tfrac{1}{2\pi}\int_{\mathbb S} d \phi \int_\Omega d \bm \omega \ {\rm Tr}\big \{\rho(\widehat H_{\bm \omega,\phi}) \partial_\phi \widehat H_{\bm \omega, \phi} \big \},
\end{equation}
with $\rho(\epsilon) = \partial_\epsilon \tilde \chi (\epsilon)$, $\int d\epsilon \, \rho(\epsilon)=1$. Recall that $\phi$ is actually a distance, hence 
\begin{equation}
\partial_\phi \widehat H_{\bm \omega, \phi}=\partial_\phi \widehat V_{\bm \omega, \phi}
\end{equation}
 is the force operator. If $\rho(\widehat H_{\bm \omega,\phi})$ is interpreted as a spectral weight, then the above formula is nothing else but the quantized averaged boundary force acting on the edge of the chain, discovered in Refs.~\cite{KellendonkJPA2004ve,JPAKellendonk2005dt}.

One question that arises is how is one going to enforce the weight distribution $\rho$ on electrons in practice? The answer is similar to that for IQHE: create an imbalance between the occupation of the edge states by applying voltage. Indeed, note that if the chain has two edges, then at equilibrium there will be boundary forces on both edges which cancel themselves exactly. But if one applies a potential bias $\Delta V$ between the two edges of a finite chain, then there will be an imbalance in the occupation of the edge states and this creates an effective weight distribution on one of the edges: 
\begin{equation}
\rho(\epsilon) = \left \{
\begin{array}{l}
1 \  {\rm if} \ \epsilon \in [\epsilon_F -\frac{1}{2}\Delta V , \epsilon_F + \frac{1}{2} \Delta V] \\
0 \ {\rm otherwise}.
\end{array}
\right .
\end{equation}
We now can state our physical prediction, namely, that there will be a net force on the chain, which, on average, is determined by the bulk invariant and by $\Delta V$:
\begin{equation}
\langle F_{\rm net} \rangle_\phi =  {\rm Ch}_1 (\bm p') \, \Delta V.
\end{equation} 
In the view of the above discussion, $\langle F_{\rm net} \rangle_\phi$ is also equal to the average force exerted spontaneously at one edge, as the edge is etched atom by atom. The potential difference $\Delta V$ can, in principle, be induced by irradiating the chain with electromagnetic waves.

\section{A Virtual Chern Insulator in 3+1-dimensions}

We now follow the program outlined in the first Chapter to generate a virtual Chern insulator in three physical dimensions and one virtual dimension. We primarily use this example to show how the strategy works in a new setting. Below we describe the steps that take us from the abstract setting all the way to the concrete physical predictions.

\subsection{The system defined}

We consider the 4-dimensional rotational algebra generated by the unitary elements $u_0, \ldots u_3$ satisfying the commutation relations:
\begin{equation}
u_i u_j = e^{i \theta_{ij}} u_j u_i, \ \theta_{ij} = - \theta_{ji} \in [0,2\pi].
\end{equation}
This algebra is generated by the monomials
\begin{equation}
u^{\bm x} = u_0^{x_0} \ldots u_3^{x_3}, \ \bm x = (x_0,\ldots,x_3) \in \mathbb Z^4,
\end{equation}
hence the elements are formal series:
\begin{equation}\label{generic}
\bm a = \sum_{\bm x \in \mathbb Z^4} \bm f_{\bm x} \, u^{\bm x},
\end{equation}
with $f_{\bm x}$ just $c$-numbers. The crossed product considered in the previous chapter is isomorphic with the algebra generated by $u_0$ and $u_1$ if we take $\theta = \theta_{01}$. In fact, the whole rotational algebra can be generated as an iterated crossed product \cite{SudoNMJ2004bc}.

The non-commutative calculus is defined by the integration:
\begin{equation}
\mathcal I (\bm a) = f_{\bm 0},
\end{equation}
and by the derivations:
\begin{equation}
\partial_j \bm a = i \sum_{\bm x \in \mathbb Z^4} x_j \, \bm f_{\bm x} \, u^{\bm x}.
\end{equation}
Given a projection $\bm p$ in this algebra, the second non-commutative Chern number is defined as \cite{ProdanJPA2013hg}:
\begin{equation}\label{Top4Chern}
{\rm Ch}_2(\bm p) = \Lambda_2 \sum_{\sigma \in S_4} (-1)^\sigma \mathcal I \big ( \bm p \prod_{j=0}^3 \partial_j \bm p \big ),
\end{equation}
where $S_4$ is the group of permutations and $(-1)^\sigma$ is the sign of the perturbation $\sigma$. $\Lambda_2$ is a proper normalization constant. It is known \cite{ProdanJPA2013hg} that ${\rm Ch}_2(\bm p)$ takes integer values which are invariant under continuous deformations of the projector ${\bm p}$. 

The $K_0$-group of this algebra is known in any dimension $d$ ($=4$ in our case), $K_0 = \mathbb Z^{2^{d-1}}$, and if all $\theta_{ij}$'s are irrational, then the 8 generators of the $K_0$ groups are known  explicitly \cite{SudoNMJ2004bc}. In particular, $K_0$ is known to contain a multitude of projections with non-zero second Chern number. Hence, the self-adjoint element:
\begin{equation}\label{GeneratorH3D}
\bm h = \sum_{j=1}^3 (u_j + u_j^\ast) + W(u_0 + u_0^\ast)
\end{equation}
will display a large number of spectral gaps and if we place the Fermi level in one of these gaps, then the Fermi projection $\bm p = \chi(\epsilon_F - \bm h)$ will have a non-zero second Chern number for many of these gaps. These affirmations can be easily tested numerically.  

We now define a representation of the algebra on the Hilbert space $\ell^2(\mathbb Z^3)$, which will lead us to the desired physical models. It is given by:
\begin{align}
u_0 & \rightarrow \pi_\phi(u_0)=e^{i (\bm \theta_0 \bm X + \phi)} \\
u_j & \rightarrow \pi_\phi(u_j) =\mathcal T_{\bm d_j}, \ j = 1,2,3,
\end{align}
where $\bm \theta_0 = (\theta_{01},\theta_{02},\theta_{03})$, $\bm X=(X_1,X_2,X_3)$ is the position operator on $\ell^2(\mathbb Z^3)$ and:
\begin{equation}
\mathcal T_{\bm d_j} |\bm y \rangle = e^{ - i \sum_{k<j} \theta_{jk} y_k} |\bm y + \bm d_j \rangle
\end{equation}
with $\bm d_j$ the generators of $\mathbb Z^3$. Let us verify the commutation relations between $u_0$ and $u_l$ for $l\geq 1$:
\begin{align}
& e^{i (\bm \theta_{0} \bm X+\phi)}\mathcal T_{\bm d_l} |\bm y\rangle 
= e^{i (\bm \theta_{0} \bm X+\phi)} e^{ - i \sum_{k<l} \theta_{lk} y_k} |\bm y + \bm d_l \rangle \nonumber \\
&= e^{i (\sum_{j=1}^3 \theta_{0j} (y_j+\delta_{jl})+\phi)} e^{ - i \sum_{k<l} \theta_{lk} y_k} |\bm y + \bm d_l \rangle,
\end{align}
and
\begin{align}
&\mathcal T_{\bm d_l} e^{i (\bm \theta_{0} \bm X + \phi)}| \bm y\rangle = \mathcal T_{\bm d_l} e^{i (\bm \theta_{0} \bm y + \phi)} |\bm y \rangle \nonumber \\
&= e^{i (\sum_{j=1}^3 \theta_{0j} y_j+\phi)} e^{ - i \sum_{k<l} \theta_{lk} y_k} |\bm y + \bm d_l \rangle.
\end{align}
Hence:
\begin{equation}
\pi_\phi(u_0) \pi_\phi(u_l) = e^{i \theta_{0l}} \pi_\phi(u_l) \pi_\phi(u_0),
\end{equation}
as required. The other commutations can be verified in the same way. The representation of a generic element Eq.~\ref{generic} is given by:
\begin{equation}\label{GenRep}
\pi_\phi(\bm a) = \sum_{\bm x \in \mathbb Z^4} f_{\bm x} e^{i (\bm \theta_0 \bm X + \phi)x_0} \mathcal T_{\bm d_1}^{x_1} \mathcal T_{\bm d_2}^{x_2} \mathcal T_{\bm d_3}^{x_3}.
\end{equation}

We now can write down our example of a Chern insulator in three physical dimensions and one virtual dimension. It is generated by $\bm h$ of Eq.~\ref{GeneratorH3D}:
\begin{equation}
H_\phi = \pi_\phi(\bm h).
\end{equation}
Explicitly:
\begin{align}\label{H3D}
H_\phi  = \sum_{j=1}^3 (\mathcal T_{\bm d_j} + \mathcal T_{\bm d_j}^\dagger)  + 2W  \cos \big (\bm \theta_{0} \bm X +\phi \big ).
\end{align}
As one can immediately see, we are talking about a crystal in a magnetic field $\bm B$ which additionally experiences an incommensurate potential in all 3-directions. The magnetic flux through the unit cell's facet in the ($jk$) planne is $\theta_{jk}$ in units $\hbar/e$. If we consider the magnetic translations on $\ell^2(\mathbb Z^3)$, written in the Landau gauge:
\begin{equation}
\mathcal T'_{\bm d_j} |\bm y \rangle = e^{ i \sum_{k>j} \theta_{jk} y_k} |\bm y + \bm d_j \rangle,
\end{equation}
then $\mathcal T'_{\bm d_j}$ and $\mathcal T_{\bm d_k}$ commute and one can see that $H_\phi$ defined in Eq.~\ref{H3D} posses the covariant property:
\begin{equation}
\mathcal T'_{\bm y} H_\phi \mathcal (T'_{\bm y})^\dagger = H_{\phi + \bm \theta_0 \bm y}, 
\end{equation}
hence it is a homogeneous system. 

The bulk invariant for this system is given by the second Chern number in Eq.~\ref{Top4Chern} applied to the Fermi projection $\bm p = \chi(\epsilon_F -\bm h)$. Our task now is to show that the bulk invariant is computable inside the physical dimensions. For this, let us write the invariant explicitly, using the physical representations $\pi_\phi(\bm p)  = \chi(\epsilon_F -H_\phi) = P_\phi$. From Eq.~\ref{GenRep}:
\begin{equation}
\mathcal I(\bm a) = f_{\bm 0} = \int_0^{2 \pi} \frac{d \phi}{2 \pi} \  \langle \bm 0 | \pi_\phi(\bm a) | \bm 0 \rangle,
\end{equation}
hence:
\begin{equation}\label{Phys4Chern}
{\rm Ch}_2(\bm p) = \Lambda_2 \sum_{\sigma \in S_4} (-1)^\sigma \int_0^{2 \pi} \frac{d \phi}{2 \pi} \ \Big \langle 0 \Big | P_\phi \prod_{j=0}^3 \partial_j P_\phi \Big | 0 \Big \rangle,
\end{equation}
where $\partial_0 P_\phi = \partial_\phi P_\phi$ and $\partial_j P_\phi = i [X_j, P_\phi]$ for $j=1,2,3$. If $\theta_{0i}$ angles are irrational, then the dynamical system $\phi \rightarrow \phi + \bm \theta_0\bm y$ is ergodic w.r.t. all three lattice translations and we can use Birkhoff's ergodic theorem
\begin{equation}
\tfrac{1}{2 \pi}\int_0^{2 \pi} d \phi f(\phi) = \lim_{V \rightarrow \infty} \frac{1}{V} \sum_{\bm y \in V} f(\phi + \bm \theta_0 \bm y)
\end{equation} 
to express the invariant as
\begin{equation}\label{Phys4Chern}
{\rm Ch}_2(\bm p) = \Lambda_2 \sum_{\sigma \in S_4} (-1)^\sigma {\rm Tr}_V\big ( P_\phi \prod_{j=0}^3 \partial_j P_\phi \Big ),
\end{equation}
where ${\rm Tr}_V$ represents the trace per volume. This shows that the invariant can be indeed computed at fixed virtual coordinate, hence our model is a virtual topological insulator.

\subsection{Physical Interpretation and Predictions}

The bulk topological invariant is connected to the isotropic part of the magneto-electric response function:
\begin{equation}
\alpha = \tfrac{1}{3} \sum_{j=1}^3 \frac{\partial P_j}{\partial B_j},
\end{equation}
where $\bm P$ is the macroscopic electric polarization vector. Indeed, according to Ref.~\cite{LeungJPA2013er} which treated the generic case with arbitrary magnetic fields and aperiodic potentials, the change in $\alpha$ when $\phi$ is varied by a whole cycle of $2\pi$ is equal to the second Chern number of the Fermi projection:
\begin{equation}
\Delta \alpha = {\rm Ch}_2(\bm p).
\end{equation}
We can write this as:
\begin{equation}
\tfrac{1}{3}\int_0^{2\pi} d \phi \  \sum_{j=1}^3 \frac{\partial^2 P_j}{\partial \phi \partial B_j} = {\rm Ch}_2(\bm p).
\end{equation}
Employing again Birkhoff's ergodic theorem and observing that the integrand is a macroscopic quantity hence invariant to lattice translations, we arrived to: 
\begin{equation}
\tfrac{1}{3} \sum_{j=1}^3 \frac{\partial^2 P_j}{\partial \phi \partial B_j}  = 2\pi \, {\rm Ch}_2(\bm p).
\end{equation}
This is our physical prediction. It can be tested experimentally in the following way. Consider the small time-modulations:
\begin{align}\label{Modulations}
& \bm B(t) = \bm B + \Delta B \cos(2 \pi f_1 t) \\
& \phi(t) = \phi + \Delta \phi \cos(2 \pi f_2 t).
\end{align}
Since the variation of $\bm P$ with time gives the electric current \cite{Schulz-BaldesCMP2013gh}:
\begin{align}
\bm J & = \partial_t \bm P,
\end{align}
one can see that the two modulations in Eq.~\ref{Modulations} will induce AC charge currents at frequencies $f_1$, $f_2$, $2f_1$, $2f_2$, $f_1 \pm f_2$ and so on, but at the frequencies $f=f_1 \pm f_2$ the amplitude of the current is:
\begin{equation}
 \bm J_{f_1 \pm f_2}  =  \pi^2 (f_1 \pm f_2) \, {\rm Ch}_2(\bm p) \Delta \phi \Delta \bm B.
\end{equation}

\subsection{Robustness Against Disorder}

Let us now consider the effect of disorder, which induces random fluctuations in the coupling constants of the  Hamiltonian:
\begin{align}\label{DisorderedH3D}
H_{\bm \omega, \phi}  & = \sum_{j=1}^3 \sum_{\bm y \in \mathbb Z^3} (1+\lambda \omega_y) |\bm y \rangle \langle \bm y | \, (\mathcal T_{\bm d_j} + \mathcal T_{\bm d_j}^\dagger) \nonumber \\
&+ 2W  \sum_{\bm y \in \mathbb Z^3}(1+\lambda' \omega'_{\bm y}) \cos  (\bm \theta_0 \bm y +\phi  )|\bm y \rangle \langle \bm y |.
\end{align}
The space $\Omega$ of the disorder configuration and the action $\tau_{\bm y}$ of $\mathbb Z^3$ on $\Omega$ can be introduced as before.  The latter is given by the shift of the disorder configuration by an $\bm y$. The disordered Hamiltonian has the covariant property with respect to the magnetic translations:
\begin{equation}
\mathcal T'_{\bm y} H_{\bm \omega,\phi} (\mathcal T'_{\bm y})^\dagger = H_{\tau_{\bm y} \bm \omega, \phi + \bm \theta_0 \bm y}.
\end{equation}

The disorder is integrated in the algebraic approach by considering the algebra generated by $C(\Omega)$ and $u_j$'s, with the additional commutation relations:
\begin{equation}
 f \, u_j = u_j \, (f \circ \tau_{\bm d_j}), \ j=1,2,3,
 \end{equation}
and $f \, u_0 = u_0 \, f $, for all $f \in C(\Omega)$. The new algebra is generated by the monomials $f (\bm \omega)\, u^{\bm x}$ and  the elements of the algebra are formal series:
\begin{equation}
\bm a = \sum_{\bm x \in \mathbb Z^4} \bm f_{\bm x} \, u^{\bm x},
\end{equation}
with $f_{\bm x}$ now complex valued functions over $\Omega$. Each element accepts a representation as an operator on the Hilbert space $\ell^2(\mathbb Z^3)$:
\begin{equation}
\pi_{\bm \omega, \phi}(\bm a) = \sum_{\bm x \in \mathbb Z^4} \sum_{\bm y \in \mathbb Z^3} f_{\bm x}(\tau_{\bm y}\bm \omega) \, |\bm y \rangle \langle \bm y | \,  \pi_\phi(u^{\bm x}),
\end{equation}
where $\pi$ is the representation already defined above. One can verify explicitly that indeed $\pi_{\bm \omega,\phi}(\bm a \bm b) = \pi_{\bm \omega,\phi}(\bm a) \pi_{\bm \omega,\phi}(\bm b)$, as required for a representation. For example, the disordered Hamiltonian in Eq.~\ref{DisorderedH3D} is generated by
\begin{equation}\label{GeneratorDisorderH3D}
\bm h' = (1+\lambda \omega_{\bm 0}) \sum_{j=1}^3 (u_j + u_j^\ast) + (1+ \lambda' \omega'_{\bm 0})W(u_0 + u_0^\ast).
\end{equation}

The integration becomes:
\begin{equation}
\mathcal I (\bm a) = \int_\Omega d\bm \omega \ f_{\bm 0}(\bm \omega),
\end{equation}
and the derivations remain unchanged. The second Chern number of a projection is formally defined as in Eq.~\ref{Top4Chern}. According to Ref.~\cite{ProdanJPA2013hg}, the second Chern number of the Fermi projection continues to take integer values that are invariant under continuous deformations of the Hamiltonians, provided the Fermi resides in a spectral gap. As in the previous Chapter, one can formulate stronger conditions that cover the regime of strong disorder. It can also be computed at fixed virtual coordinate and from a single disorder configuration.

The non-commutative second-Chern number can be evaluated numerically using the methods developed in Refs.~\cite{ProdanAMRX2013bn}. Explicit numerical calculations can be found in Leung's PhD thesis \cite{LeungPhDThesis2013hd}.

\subsection{The Half-Space Algebra}

Let us now constrain the model to half of the space, $x_3 \geq 0$. Following the arguments from the previous chapter, we define the half-space algebra as the algebra generated by $C(\Omega)$ and $\hat u_0, \ldots, \hat u_3$ satisfying the same commutation relations as before, with the exception that $\hat u_3$ is modified into a partial isometry:
\begin{equation}\label{Isometry2}
\hat u_3^\ast \hat u_3 = 1, \ \ \hat u_3 \hat u_3^\ast = 1- \hat e_3.
\end{equation}
The half-space algebra is generated by the monomials:
\begin{equation}
f(\omega) \hat u_0^{x_0} \hat u_1^{x_1} \hat u_2^{x_2} \hat u_3^m (\hat u_3^\ast)^n = f(\omega) \hat u^{\hat{\bm x} }\hat u_3^m (\hat u_3^\ast)^n,
\end{equation}
with $\hat{\bm x} = (x_0,x_1,x_2) \in \mathbb Z^3.$ Hence the elements are formal series:
\begin{equation}
\hat{\bm a} = \sum_{m,n \in \mathbb N} \sum_{\hat{\bm x} \in \mathbb Z^3} f_{mn}(\bm \omega,\hat {\bm x}) \hat u ^{\hat{\bm x}}\hat u_3^m (\hat u_3^\ast)^n.
\end{equation}

Any element $\bm a = \sum_{\bm x\in \mathbb Z^4 } f_{\bm x}(\omega) u^{\bm x}$ from the bulk algebra can be transformed into an element from the half-space algebra:
\begin{equation}
\bm a \rightarrow j(\bm a) = \sum_{(\hat{\bm x},x_3)\in \mathbb Z^4} f_{\hat{\bm x},x_3} \hat u^{\hat{\bm x}}  \hat u_3^{x_3},
\end{equation}
with $\hat u_3^{-1}$ interpreted as $\hat u_3^\ast$, and reciprocally, any element 
\begin{equation}
\hat{\bm a} = \sum_{m,n\in \mathbb N} f_{mn}(\omega,\hat{\bm x}) \hat u^{\hat{\bm x}} \hat u_3^m (\hat u_3^\ast)^n
\end{equation} 
from the half-space algebra can be transformed into an element from the bulk-algebra:
\begin{equation}
\hat{\bm a} \rightarrow {\rm ev}(\hat{\bm a}) = \sum_{m,n\in \mathbb N}\sum_{\hat{\bm x} \in \mathbb Z^3} f_{mn}(\omega,\hat{\bm x})  u^{\hat{\bm x}}  u_3^{m-n}.
\end{equation}
As before, any element of the half-space algebra can be written uniquely as:
\begin{equation}
\hat{\bm a} = j(\bm a) + \tilde{\bm a},
\end{equation}
with
\begin{equation}
\bm a = {\rm ev}(\hat{\bm a}) \ \ {\rm and} \ \ \tilde{\bm a} = \hat{\bm a}- j(\bm a).
\end{equation}
The component $\tilde{\bm a}$ is necessarily of the form:
\begin{equation}
\tilde{\bm a} = \sum_{m,n\in \mathbb N} \sum_{\hat{\bm x} \in \mathbb Z^3}\tilde f_{mn}(\bm \omega,\hat{\bm x}) \hat u^{\hat{\bm x}} \hat u_3^m \, \hat e_3 \, (\hat u_3^\ast)^n,
\end{equation}
and the elements like $\tilde{\bm a}$ form an ideal inside the half-space algebra, which we call the boundary-algebra. 

The elements of the half-space algebra accepts a canonical representation as operators over $\ell^2(\mathbb Z^2 \times \mathbb N)$, given by:
\begin{equation}
\hat \pi_{\bm \omega,\phi}(f) = \Pi \big (\sum_{\bm y \in \mathbb Z^3} f(\tau_{\bm y} \bm \omega) |{\bm y}\rangle \langle \bm y| \big ) \Pi 
\end{equation}
and:
\begin{align}
&\hat \pi_{\bm \omega,\phi}(\hat u_0)=e^{i (\bm \theta_0 \bm X + \phi)} \Pi \\
&\hat \pi_{\bm \omega,\phi}(\hat u_j) =\mathcal T_{\bm d_j} \Pi, \ j = 1,2,3,
\end{align}
with $\Pi$ the projection from $\ell^2(\mathbb Z^3)$ to $\ell^2(\mathbb Z^2 \times \mathbb N)$. The family of half-space Hamiltonians obtained from the bulk $H_{\bm \omega,\phi}$, via open boundary conditions, $\Pi H_{\bm \omega,\phi} \Pi$ are generated by:
\begin{equation}
j(\bm h') =(1+\lambda \omega_{\bm 0}) \sum_{j=1}^3 (\hat u_j + \hat u_j^\ast) + (1+ \lambda' \omega'_{\bm 0})W(\hat u_0 + \hat u_0^\ast).
\end{equation}
The open boundary condition can be changed to any other boundary condition by adding an element from the edge algebra:
\begin{equation}
\tilde{\bm h}'= \sum_{m,n\in \mathbb N} \tilde h_{mn}(\bm \omega, \hat{\bm x}) \hat u^{\hat{\bm x}} \, \hat u_3^m \, \hat e \, (\hat u_3^\ast)^n.
\end{equation}
This term can account for the lattice distortions or relaxations near the edge, chemical contamination, bond breaking and so on.  

\subsection{The Boundary Invariant}

We need to define a non-commutative differential calculus over the edge algebra in order to define the boundary invariant. This is given by the integration:
\begin{equation}
\widetilde{I}(\tilde{\bm a}) = \sum_{n \in \mathbb N} \int_\Omega d \bm \omega \ \tilde f_{nn}(\bm \omega, 0),
\end{equation}
which is well defined only if $\tilde f_{nm}$ is localized near the boundary, and by the derivations:
\begin{align}
\tilde \partial_j \tilde{\bm a} = i \sum_{m,n\in \mathbb N} \sum_{\hat{\bm x} \in \mathbb Z^3} x_j \, \bm f_{mn}(\bm \omega, \hat{\bm x}) \,  \hat u^{\hat{\bm x}} \, \hat u_3^m \, \hat e \, (\hat u_3^\ast)^n,
\end{align}
for $j=0,1,2$.

We now restrict the bulk Hamiltonian to half of the space by imposing a generic boundary condition through an element $\tilde{\bm h}'$ from the boundary algebra:
\begin{equation}
\hat{\bm h}' = j(\bm h') + \tilde{\bm h}'.
\end{equation}
As before, we define the element:
\begin{equation}
\hat{\bm g}' = \tilde \chi(\epsilon_f - \hat{\bm h}')
\end{equation}
such that ${\rm ev}(\hat{\bm g}') = \bm p'$, the Fermi projector of $\bm h'$, and then the unitary element
\begin{equation}
\hat{\bm u}' = \exp ( -2 \pi i \hat{\bm g}').
\end{equation}
The boundary invariant is then defined as the 3-dimensional winding number of this unitary element:
\begin{equation}\label{3DWinding}
\nu_3( \hat{\bm u}') = \Lambda_3 \sum_{\sigma \in S_3} (-1)^\sigma \, \widetilde{I} \, \big( \prod_{j=0}^2 (\hat{\bm u}'-1) \tilde \partial_j (\hat{\bm u}' -1) \big ),
\end{equation}
with $\Lambda_3$ a proper normalization constant. According to Ref.~\cite{ProdanOddChernArxiv2014}, the non-commutative odd Chern number of unitary elements takes integer values and is invariant to continuous deformations. And according to Ref.~\cite{Kellendonk:2002of} (see Theorem A10):
\begin{equation}\label{BulkBoundary3D}
\nu_3( \hat{\bm u}') = {\rm Ch}_2(\bm p').
\end{equation} 
As in $1+1$ dimension, the boundary invariant can be computed at fixed virtual coordinate by cleaving the surface layer by layer and averaging the quantity inside the formal integral in Eq.~\ref{3DWinding}.

To determine the physical meaning of this invariant one needs a formula like Eq.~\ref{EdgeInvariant3}, which is not available yet. Nevertheless, we point out that it is this boundary invariant that assures the existence of boundary states of the half-space Hamiltonian:
\begin{equation}
H_{\bm \omega,\phi} = \hat{\pi}_{\bm \omega, \phi}(\hat{\bm h}),
\end{equation}
when $\phi$ is scanned over the interval $[0,2\pi]$. Indeed, the spectrum of $\hat h$ is the reunion of all spectra of $H_{\bm \omega,\phi}$ and if this spectrum is gapped, then we can deform $\tilde \chi$ such that its variation occurs entirely inside this gap. Consequently, $\hat{\bm u}_F-1 = 0$ and the boundary invariant is zero. Then equality \ref{BulkBoundary3D} ensures that, whenever the bulk invariant is no zero, the bulk gap is entirely filled with boundary spectrum. 

\section{Conclusions}

Our conclusion is that Refs.~\cite{KrausPRL2012hh,KrausPRL2013fj} have indeed introduced a new class of strong topological systems. We are particularly excited about it because we are no longer bound to the 3 physical dimensions, which will certainly lead to the discovery of many new strong topological insulators. It will be interesting to explore the other symmetry classes and see exactly how the virtual topological insulators are constructed there.

\section*{ACKNOWLEDGMENTS}
We acknowledge support from the U.S. NSF grant DMR-1056168.

\bibliography{../../../../TopologicalInsulators}

\end{document}